\documentclass[11pt]{article}

\usepackage[utf8]{inputenc}
\usepackage{mathpazo}
\usepackage{bm}
\usepackage{aurical}
\usepackage{amsmath, amssymb}
\usepackage{import}
\usepackage[toc,page]{appendix}
\usepackage{dsfont}
\usepackage{mathtools}
\usepackage{float}
\usepackage{graphicx}
\usepackage{cite}
\usepackage{hyperref}
\usepackage{setspace}
\usepackage{color}
\usepackage{slashed}
\usepackage{cleveref}
\numberwithin{equation}{section}
\usepackage[colorinlistoftodos]{todonotes}
\usepackage[affil-it]{authblk}
\usepackage{float}
\usepackage{indentfirst}
\usepackage{soul}
\usepackage{booktabs}
\usepackage{eso-pic,graphicx}
\usepackage{nicefrac}
\usepackage{epsfig}
\usepackage{amsfonts,mathtools,amssymb,siunitx,mathrsfs}
\usepackage{graphicx, nicefrac}
\usepackage[numbers,sort&compress]{natbib}
\newcommand{\str}{\mathrm{Str}}
\newcommand{\e}{\mathrm{e}}
\newcommand{\jsn}{\mathrm{sn}}
\newcommand{\jcn}{\mathrm{cn}}
\newcommand{\mi}{\mathrm{i}\,}
\usepackage{braket}

\usepackage{geometry}
\usepackage{color,hyperref}
\definecolor{darkblue}{rgb}{0.0,0.0,0.3}
\hypersetup{colorlinks,breaklinks,
            linkcolor=darkblue,urlcolor=darkblue,
            anchorcolor=darkblue,citecolor=darkblue}

\title{The tetrahedral Zamolodchikov algebra for the fermionic Bazhanov-Stroganov $R$-operator}
\author[1]{A. Melikyan\footnote{\href{mailto:amelik@gmail.com}{amelik@gmail.com}}}
\affil[1]{Instituto de Física\\
Universidade de Brasília\\
70910-900, Brasília, DF, Brasil}

\usepackage{palatino}

\begin{document}
\maketitle

\begin{abstract}
We find the fermionic R-operator based on Bazhanov-Stroganov three-parameter elliptic parametrization of the free fermion model, and the corresponding Yang-Baxter and decorated Yang-Baxter equations, which are of the difference type in one of the spectral parameters. We also find a solution of the corresponding tetrahedral Zamolodchikov algebra for a specific case.
\end{abstract}
\section{Introduction}
The integrable properties of fermionic $(1+1)$-dimensional models have attracted much attention due to their explicit appearance in string theory (see \cite{Beisert:2010jr} for a review), and, on the other hand, their relation to the one-dimensional Hubbard model \cite{Essler:2005bk}, the $R$-matrix of which  has been proposed to be related to the $S$-matrix of strings via $AdS/CFT$ correspondence \cite{Mitev2017}. More specifically, the following fermionic model arises  from the string theory  on $AdS_5 \times S^5$ when restricted to the $su(1 \vert 1)$  subsector \cite{Arutyunov:2004yx}:
\begin{align}
	 \mathscr{L}&= \mi \bar{\psi} \gamma_{\mu} \partial^{\mu} \psi \: - m \bar{\psi} \psi + \frac{g_2}{4m} \epsilon^{\alpha \beta} \left( \bar{\psi}
	\partial_{\alpha} \psi \; \bar{\psi}\: \gamma^3 
	\partial_{\beta} \psi -
	\partial_{\alpha}\bar{\psi} \psi \; 
	\partial_{\beta} \bar{\psi}\: \gamma^3 \psi \right)-  \frac{g_3}{16m} \epsilon^{\alpha \beta} \left(\bar{\psi}\psi\right)^2 
	\partial_{\alpha}\bar{\psi}\:\gamma^3
	\partial_{\beta}\psi. \label{intro:aaf_lag}
\end{align}
It can be shown to be a completely  integrable classically, and its various quantum integrability properties has been investigated from various points of view \cite{Melikyan:2011uf,Melikyan:2012kj,Melikyan:2014yma,Melikyan2019ac,Melikyan:2016gkd,Melikyan:2014mfa}. Despite this, the quantization of this model by direct methods of integrable systems was not successful as a consequence of the non-ultralocal nature of the algebra of Lax operators, given in \cite{Melikyan:2012kj,Melikyan:2014yma}. Quantization of such non-ultralocal  models remains an open problem, and their essential features can already be captured by considering the free fermionic model. The Lax pair for the latter can be obtained from that of the full model \eqref{intro:aaf_lag} given in \cite{Melikyan:2012kj}, by setting the coupling constants $g_2$ and $g_3$ to zero, and has the form:
\begin{align}
		 L_{0}(x;\mu) &= {\xi}_{0}^{\scriptscriptstyle{(\tau)}}(x;\mu) \mathbb{1} + {\xi}_{1}^{\scriptscriptstyle{(\tau)}}(x;\mu) \sigma^{3} + {\Lambda}^{\scriptscriptstyle{(-)}}_{\tau}(x;\mu) \sigma^+ + {\Lambda}^{\scriptscriptstyle{(+)}}_{\tau}(x;\mu) \sigma^-, \label{intro:Lax_L0} \\
	 L_{1}(x;\mu) &= {\xi}_{0}^{\scriptscriptstyle{(\sigma)}}(x;\mu) \mathbb{1} + {\xi}_{1}^{\scriptscriptstyle{(\sigma)}}(x;\mu) \sigma^{3} + {\Lambda}^{\scriptscriptstyle{(-)}}_{\sigma}(x;\mu) \sigma^+ + {\Lambda}^{\scriptscriptstyle{(+)}}_{\sigma}(x;\mu) \sigma^- \label{intro:Lax_L1},
\end{align}
where we have denoted $\chi_1 = \nicefrac{(\psi_1 - \psi_2)}{\sqrt{2}}, \chi_2 = \mi \nicefrac{(\psi_1+ \psi_2)}{\sqrt{2}}, \chi_{3}\equiv\chi_{1}^{*}, \chi_{4}\equiv\chi_{2}^{*}$;	$l_0 = 1,  l_1 = \cosh(2\mu), l_2 = -\sinh(2\mu),  l_3 = \cosh (\mu),  l_4 = \sinh(\mu)$;  and the functions ${\xi}_{j}^{\scriptscriptstyle{(\sigma,\tau)}}(x;\mu)$, and $t{\Lambda}_{\sigma,\tau}^{\scriptscriptstyle{(\pm)}}(x;\mu)$ have the following explicit form:
\begin{align}
	{\xi}^{\scriptscriptstyle{(\sigma)}}_0 &= \frac{1}{4} \left[ - \chi_3\chi_1' + \chi_4\chi_2' - \chi_1\chi_3' + \chi_2\chi_4' \right], \quad
	{\xi}^{\scriptscriptstyle{(\sigma)}}_1 = \frac{\mi l_2 }{2}, \label{free:xis1} \\
{\Lambda}^{\scriptscriptstyle{(-)}}_{\sigma} &=  \left[ -l_{3}\chi_2'- il_{4}\chi_1' \right], \quad
	{\Lambda}^{\scriptscriptstyle{(+)}}_{\sigma} =\left[ -l_{3}\chi_4'+ il_{4}\chi_3' \right],\label{free_app:Lambdasp}\\
{\xi}^{\scriptscriptstyle{(\tau)}}_0 &= \frac{i}{2} \left[\chi_{3}\chi_{1}+\chi_{4}\chi_{2} \right]+ \frac{1}{4}\left[-\chi_{3}\dot{\chi}_{1}-\chi_{1}\dot{\chi}_{3} +\chi_{4}\dot{\chi}_{2}  +\chi_{2}\dot{\chi}_{4} \right],  \quad
	{\xi}^{\scriptscriptstyle{(\tau)}}_1 = -\frac{\mi l_{1}}{2}, \label{free_app:ksi1tau}\\
	{\Lambda}^{\scriptscriptstyle{(-)}}_{\tau} &= -i \left[l_{3}\chi_{2} - \mi l_{4}\chi_{1} \right] -  \left[l_{3}\dot{\chi}_{2} + il_{4}\dot{\chi}_{1}  \right], \quad
	{\Lambda}^{\scriptscriptstyle{(+)}}_{\tau} = i \left[l_{3}\chi_{4} + \mi l_{4}\chi_{3} \right] -  \left[l_{3} \dot{\chi}_{4} - il_{4}\dot{\chi}_{3}  \right].\label{free_app:Lambdataup}
\end{align}

The non-ultralocality nature of the theory is revealed by the presence of nonzero $B(x,y;\lambda,\mu)$ and $C(x,y;\lambda,\mu)$ coefficients in the algebra:
\begin{align}
	\{ {L}_{1}(x;\lambda) \overset{\otimes}{,} {L}_{1}(y;\mu) \} &= A(x,y;\lambda,\mu) \delta(x-y) +  B(x,y;\lambda,\mu) \partial_{x}\delta(x-y)+C(x,y;\lambda,\mu) \partial^{2}_{x}\delta(x-y), \label{intro:nonultralocal}
\end{align}
where the explicit form of the coefficients $A(x,y;\lambda,\mu),B(x,y;\lambda,\mu)$ and $C(x,y;\lambda,\mu)$ can be found in \cite{Melikyan:2012kj,Melikyan:2014yma}. We stress that the above algebraic structure remains the same for the full model interacting model \eqref{intro:aaf_lag}. Moreover, the full superstring theory on on $AdS_5 \times S^5$ can also be shown to be of this type \cite{Das:2004hy,Das:2005hp}. The presence of non-ultralocal terms in  the algebra does not allow obtaining a well-defined algebraic structure of the monodromies, and, as a consequence, it is not possible to use the standard techniques of quantization by formulating the lattice version of the model and by means of the Bethe ansatz.

One can, however, start with the lattice formulation of a model, and trace the appearance of the non-ultralocal terms in the continuous limit.\footnote{The results of this investigation will be presented in the upcoming publication.} While the lattice formulation of the  model in \eqref{intro:aaf_lag} is not known, that of the free fermion model is indeed well-known, and since the non-ultralocal algebraic structure \eqref{intro:nonultralocal} is the same for both models, one hopes to gain important insights towards the goal of quantization of the full model by restricting to the simpler free fermion model. Moreover, quite surprisingly, the $S$-matrix of the full string theory appears to be related to the $R$-matrix of the one-dimensional Hubbard model \cite{Essler:2005bk,Mitev2017}, which itself can be obtained, as shown by Shastry, by pairing two $R$-matrices corresponding to the free fermion model. The construction however involves $R$-matrices which do not depend on the differences of the spectral parameters. This is due to the so-called decorated Yang-Baxter equation, where the $R_{jk}$-matrix dependence is not of the difference type $(u_j-u_k)$, unlike that of the Yang-Baxter equation. Thus, its not obvious how to obtain, in the continuous limit, the $(1+1)$-relativistic fermion model as, for example, appearing from string theory \eqref{intro:aaf_lag}.

There exists, however, a more general three-parameter parametrization of the free fermion model due to Bazhanov and Stroganov \cite{Bazhanov:1984iw,Bazhanov:1984ji,Bazhanov:1984jg}, which as we show in this Letter, allows a formulation of both  Yang-Baxter and decorated  Yang-Baxter relations in the form where the dependence of  $R_{jk}$-matrix is indeed of the difference type with respect to one of the spectral parameters. To this end, we  use a more convenient for our purposes fermionic $R$-operator formalism given in \cite{Umeno1998b,Umeno1998,Umeno2000}, and the present the fermionic versions of the Yang-Baxter and decorated  Yang-Baxter equations. Furthermore, towards the goal of constructing an interacting theory, we find a solution for the corresponding to our $R$-operator tetrahedral Zamolodchikov algebra \cite{Zamolodchikov1981,Korepanov1993,Korepanov2013,Korepanov1994c} for a specific choice of parameters.

\section{Bazhanov-Stroganov's $R$-matrix}
\subsection{Free fermion model and elliptic parametrization}
Bazhanov and Stroganov had investigated in \cite{Bazhanov:1984iw,Bazhanov:1984ji,Bazhanov:1984jg} an interesting general solution of the inhomogeneous eight-vertex model with the  $R$-matrix of the form: 
\begin{align}
    \hat{R}=\begin{pmatrix}
a & 0 & 0 & d\\
0 & b & c & 0 \\
0 & c' & b' & 0 \\
d' & 0 & 0 & a \label{bs:R_matrix_orig},
\end{pmatrix},
\end{align}
satisfying the free fermion condition \cite{Fanwu723}:
\begin{align}
    a a' + b b' - c c' - d d'=0 \label{bs:free_fermion_condition}
\end{align}
The coefficients in \eqref{bs:R_matrix_orig} are parametrized by the spectral parameter $u$ and two additional rapidities $\zeta_{1}$ and $\zeta_{2}$:
\begin{align}
    &a(u;\zeta_{1},\zeta_{2}) =\rho \left[ 1-\e(u)\e(\zeta_{1})\e(\zeta_{2}) \right]; \quad a'(u;\zeta_{1},\zeta_{2}) =\rho \left[ \e(u)-\e(\zeta_{1})\e(\zeta_{2}) \right] \label{bs:a_a_prime} \\
    &b(u;\zeta_{1},\zeta_{2}) =\rho \left[ \e(\zeta_{1})-\e(u)\e(\zeta_{2}) \right]; \quad b'(u;\zeta_{1},\zeta_{2}) =\rho \left[ \e(\zeta_{2})-\e(u)\e(\zeta_{1}) \right] \label{bs:b_b_prime}\\
    &c(u;\zeta_{1},\zeta_{2})=c'(u;\zeta_{1},\zeta_{2}) =\rho \: \jsn^{-1}\left(\frac{u}{2}\right)\left[ 1-\e(u)\right]\left[\e(\zeta_{1})e(\zeta_{2})\jsn(\zeta_{1})\jsn(\zeta_{2}) \right]^{1/2}, \label{bs:c_c_prime}\\
    &d(u;\zeta_{1},\zeta_{2})=d'(u;\zeta_{1},\zeta_{2}) =- \mi k \rho \: \jsn \left(\frac{u}{2} \right)\left[ 1+\e(u)\right]\left[\e(\zeta_{1})e(\zeta_{2})\jsn(\zeta_{1})\jsn(\zeta_{2}) \right]^{1/2} \label{bs:d_d_prime}.
\end{align}
Here the functions $\jsn(x)$ and $\jcn(x)$ are the Jacobi elliptic functions of modulus $k$ \cite{whittaker_watson_1996}, $\e(x)$ is the elliptic exponential $\e(x)=\jcn(x) + \mi \jsn(x)$, and $\rho$ is an arbitrary factor.
The $R$-matrix in \eqref{bs:R_matrix_orig} satisfies the Yang-Baxter equation:
\begin{align}
   \hat{R}_{12}(\eta_{12};\zeta_{1},\zeta_{2})\hat{R}_{13}(\eta_{13};\zeta_{1},\zeta_{3})\hat{R}_{23}(\eta_{23};\zeta_{2},\zeta_{3})=\hat{R}_{23}\eta_{23};\zeta_{2},\zeta_{3})\hat{R}_{13}(\eta_{13};\zeta_{1},\zeta_{3})\hat{R}_{12}(\eta_{12};\zeta_{1},\zeta_{2}),\label{bs:YBE}
\end{align}
where we have used the shorthand notation $\eta_{jk} \equiv u_{j}-u_{k}$. 

\subsection{Fermionic $R$-operator}

We now introduce,  following \cite{Umeno1998b,Umeno1998}, the fermionic $R$-operator corresponding to the $R$-matrix \eqref{bs:R_matrix_orig}. This fermionic versions of the $R$-matrix and the Yang-Baxter equations are convenient to use from the beginning in order to make a connection with the Lax connections in \eqref{intro:Lax_L0} and \eqref{intro:Lax_L1}, which are written in terms of fermionic variables. To do so one has to apply the Jordan-Wigner transformation (see \cite{Essler:2005bk} for an extensive treatment) to the above $\hat{R}$-matrix as well as the the Yang-Baxter equation \eqref{bs:YBE}. The essential technical details are explained in \cite{Umeno1998b,Umeno1998} and are omitted here. The final result when applied to the  case with the $R$-matrix \eqref{bs:R_matrix_orig} is as follows: First, the associated fermionic $R$-operator, written in terms of the fermionic variables $(c^{\dagger}_{k},c_{k});$ $\{c^{\dagger}_{j},c_{k}\}=\delta_{jk}$, takes the from:
\begin{align}
    R_{jk}(u;\zeta_{j},\zeta_{k})&=a(u;\zeta_{j},\zeta_{k})\left[-n_{j} n_{k} \right] +a(u;\zeta_{j},\zeta_{k})\left[(1-n_{j})(1- n_{k}) \right]+b(u;\zeta_{j},\zeta_{k})\left[n_{j}(1- n_{k}) \right] \nonumber\\
    &+b'(u;\zeta_{j},\zeta_{k})\left[n_{k}(1- n_{j})\right] +c(u;\zeta_{j},\zeta_{k})\left[\Delta_{jk}+\Delta_{kj} \right] + d(u;\zeta_{j},\zeta_{k})\left[-\tilde{\Delta}^{(\dagger)}_{jk}-\tilde{\Delta}_{jk}\right] ,\label{bs:fermionic_R}
\end{align}
where  $n_{k}=c^{\dagger}_{k}c_{k}$, $\Delta_{jk}=c^{\dagger}_{j}c_{k},\tilde{\Delta}^{(\dagger)}_{jk}=c^{\dagger}_{j}c^{\dagger}_{k}$ and $\tilde{\Delta}_{jk}=c_{j}c_{k}$. Furthermore, it can be shown that the fermionic $R$-operator \eqref{bs:fermionic_R} satisfies the Yang-Baxter equation:
\begin{align}
    R_{12}(\eta_{12};\zeta_{1},\zeta_{2})R_{13}(\eta_{13};\zeta_{1},\zeta_{3})R_{23}(\eta_{23};\zeta_{2},\zeta_{3})=R_{23}(\eta_{23};\zeta_{2},\zeta_{3})R_{13}(\eta_{13};\zeta_{1},\zeta_{3})R_{12}(\eta_{12};\zeta_{1},\zeta_{2}),\label{bs:YBE_fermionic}
\end{align}
The corresponding fermionic monodromy operator is defined in the usual manner:
\begin{align}
    T_{a}(u;\{\zeta_{j}\};\zeta_{a})=R_{aN}(u;\zeta_{a},\zeta_{N})R_{a,N-1}(u;\zeta_{a},\zeta_{N-1})\cdot\ldots \cdot R_{a1}(u;\zeta_{a},\zeta_{1})\label{bs:monodromy}
\end{align}
which satisfies the $RTT=TTR$ relation:
\begin{align}
     R_{ab}(u-v;\zeta_{a},\zeta_{b}) T_{a}(u;\{\zeta_{j}\};\zeta_{a}) T_{b}(v;\{\zeta_{j}\};\zeta_{b})=T_{b}(v;\{\zeta_{j}\};\zeta_{b})T_{a}(u;\{\zeta_{j}\};\zeta_{a})R_{ab}(u-v;\zeta_{a},\zeta_{b}). \label{bs:rtt_ttr}
\end{align}
Defining also:\footnote{For the definition of the supertrace $\str_{a}\left[F\right]$ over the auxiliary space $a$ see \cite{Umeno1998b}}:
\begin{align}
    \tau(u;\{\zeta_{j}\};\zeta_{a})=\str_{a}\left[ T_{a}(u;\{\zeta_{j}\};\zeta_{a})\right], \label{bs:tau}
\end{align}
one obtains the commuting quantities:
\begin{align}
    \left[\tau(u;\{\zeta_{j}\};\zeta_{a}),\tau(v;\{\zeta_{j}\};\zeta_{b}) \right]=0.
\end{align}

As an aplication, we apply the fermionic Yang-Baxter relation \eqref{bs:YBE_fermionic} for the case  of equal parameters $\zeta_{i} \equiv \zeta$, and compute the Hamiltonian:
\begin{align}
    \hat{\mathcal{H}}=\tau^{-1}(0;\zeta)\frac{d}{du}\tau(u;\zeta)\vert_{u=0}. 
\end{align}
Using the explicit form of the coefficients \eqref{bs:a_a_prime}-\eqref{bs:d_d_prime}, and the relations:
\begin{align}
    R_{jk}(0;\zeta)&=\beta \mathcal{P}_{jk}\\
    \tau(0;\zeta)&=\beta^{N}\mathcal{P}_{12}\mathcal{P}_{23}\cdot \ldots \cdot \mathcal{P}_{N,N-1},
\end{align}
where we denoted $\beta=(-2 \mi \rho)\e(\zeta) \jsn(\zeta)$, and  $\mathcal{P}=1-n_{j}-n_{k}+\Delta_{jk}+\Delta_{kj}$ is the permutation operator, one immediately obtains the Hamiltonian for the fermionic $XY$ model in the external field:\footnote{To compare, the construction in  \cite{Umeno1998b,Umeno1998} is rather non-straightforward and more involved, requiring also the decorated Yang-Baxter relation. We also note that the parameter $\zeta$ corresponds to the external field.}
\begin{align}
   \hat{\mathcal{H}}= \frac{1}{2\jsn(\zeta)}\sum_{j=1}^{N}\left[\left(\Delta_{j,j+1}+\Delta_{j+1,j}\right)+ k\jsn(\zeta) \left(\tilde{\Delta}^{(\dagger)}_{j,j+1}-\tilde{\Delta}_{j+1,j}\right)+2\jcn(\zeta)\left(n_{j}-\nicefrac{1}{2}\right)\right]. \label{bs:XY_Hamiltonian}
\end{align}

To obtain the Lax connections \eqref{intro:Lax_L0}, \eqref{intro:Lax_L1} and the Lagrangian \eqref{intro:aaf_lag} from the free fermion model described by the fermionic $R$-operator \eqref{bs:fermionic_R} one has to extend the above construction for the spinless fermions to include two copies of the $R^{(s)}$-operator for each spin $s=\ket{\uparrow},\ket{\downarrow}$. The generalization is straightforward, and the free fermion model with spin $s=\nicefrac{1}{2}$ is obtained from two copies of the fermionic $R$-matrix \eqref{bs:fermionic_R}:
\begin{align}
   \mathcal{R}_{jk}(u_{j}-u_{k};\zeta_{j},\zeta_{k}):=R^{(\uparrow)}_{jk}(u_{j}-u_{k};\zeta_{j},\zeta_{k})R^{(\downarrow)}_{jk}(u_{j}-u_{k};\zeta_{j},\zeta_{k}). \label{bs:fermionic_R_spin}
\end{align}
The fermionic operator $\mathcal{R}_{jk}(u;\zeta_{j},\zeta_{k})$ in \eqref{bs:fermionic_R_spin} satisfies the same Yang-Baxter equation \eqref{bs:YBE_fermionic}, and the corresponding monodromy operator and commuting quantities can be constructed in the same manner as above. Starting from the $\mathcal{R}_{jk}(u;\zeta_{j},\zeta_{k})$, corresponding to the case considered above of equal parameters $\zeta_{i} \equiv \zeta$, one obtains two non-interacting fermionic $XY$ models in external fields, with the $\mathcal{R}$-matrix \eqref{bs:fermionic_R_spin} which is of the difference type in the spectral parameter, unlike the case in \cite{Umeno1998b,Umeno1998}. The details of obtaining the corresponding Lax pair and passage to the continuous limit will be presented in the upcoming publication.

\section{Tetrahedral Zamolodchikov algebra}
\subsection{Decorated Yang-Baxter equation}

We now turn our attention to the interacting case. First, we derive the so-called decorated Yang-Baxter equation. It follows from the following relation:
\begin{align}
    (2n_{j,(s)}-1)R^{(s)}_{jl}(u;\zeta_{j},\zeta_{l};k) (2n_{l,(s)}-1)=R^{(s)}_{jl}(u;\zeta_{j}+2 \mathrm{K}(k),\zeta_{l}+2 \mathrm{K}(k);-k);\quad s=\ket{\uparrow},\ket{\downarrow} \label{tza:identity_main}
\end{align}
where we have written the dependence on the modulus $k$ in $R^{(s)}_{jl}(u;\zeta_{j},\zeta_{k};k)$ explicitly, and $\mathrm{K}(k)$ is the complete elliptic integral of the first kind \cite{whittaker_watson_1996}. The above formula can be checked using the formulas given in the Appendix, as well as  the explicit expressions for the functions \eqref{bs:a_a_prime}-\eqref{bs:d_d_prime}. Then, the decorated Yang-Baxter equation is a relation that is derived from the Yang-Baxter equation \eqref{bs:YBE_fermionic}, together with the above identity \eqref{tza:identity_main}. It has the form:
\begin{align}
    R^{(s)}_{12}(\eta_{12};\zeta_{1},\zeta_{2}-2\mathrm{K}(k);k) (2n_{1,(s)}-1) R^{(s)}_{13}(\eta_{13};\zeta_{1},\zeta_{3}-2\mathrm{K}(k);-k) R^{(s)}_{23}(\eta_{23};\zeta_{2},\zeta_{3};k) \nonumber \\
    =R^{(s)}_{23}(\eta_{23};\zeta_{2},\zeta_{3};k)R^{(s)}_{13}(\eta_{13};\zeta_{1},\zeta_{3}-2\mathrm{K}(k);-k)(2n_{1,(s)}-1)R^{(s)}_{12}(\eta_{12};\zeta_{1},\zeta_{2}-2\mathrm{K}(k);k) \label{tza:DYBE}
\end{align}
Unlike the previously considered cases \cite{Umeno1998b}, the decorated Yang-Baxter relation \eqref{tza:DYBE} depends on the differences of the spectral parameters  $\eta_{jk} \equiv u_{j}-u_{k}$, taking an asymmetrical form with respect to the other parameters $\zeta_{i}$. Using \eqref{tza:identity_main} and taking the product of two copies of \eqref{tza:DYBE} for $s=\ket{\uparrow},\ket{\downarrow}$ one can readily arrive at a similar identity for $\mathcal{R}_{jk}(u;\zeta_{j},\zeta_{k})$ defined in \eqref{bs:fermionic_R_spin}.

Using the notations:
\begin{align}
    \mathrm{L}^{0,(s)}_{jk} &=R^{(s)}_{jk}(\eta_{jk};\zeta_{j},\zeta_{k};k) \label{tza:L0}\\
    \mathrm{L}^{1,(s)}_{jk} &=R^{(s)}_{jk}(\eta_{jk};\zeta_{j},\zeta_{k}-2\mathrm{K}(k);-k)(2n_{j,s}-1), \label{tza:L1}
\end{align}
we write the Yang-Baxter and decorated Yang-Baxter equations in the form:
\begin{align}
    \mathrm{L}^{0,(s)}_{12} \mathrm{L}^{0,(s)}_{13} \mathrm{L}^{0,(s)}_{23} &=  \mathrm{L}^{0,(s)}_{23} \mathrm{L}^{0,(s)}_{13} \mathrm{L}^{0,(s)}_{12}, \label{tza:L000} \\
     \mathrm{L}^{0,(s)}_{12} \mathrm{L}^{1,(s)}_{13} \mathrm{L}^{1,(s)}_{23} &=  \mathrm{L}^{1,(s)}_{23} \mathrm{L}^{1,(s)}_{13} \mathrm{L}^{0,(s)}_{12}. \label{tza:L011}
\end{align}
The tetrahedral Zamolodchikov algebra is an algebraic expression of the form:
\begin{align}
    \mathrm{L}^{\alpha_{1},(s)}_{12} \mathrm{L}^{\alpha_{2},(s)}_{13} \mathrm{L}^{\alpha_{3},(s)}_{23} =\sum_{\beta_{i}=0,1} S^{\alpha_{1} \alpha_{2} \alpha_{3}}_{\beta_{1}\beta_{2}\beta_{3}}\mathrm{L}^{\beta_{1},(s)}_{23} \mathrm{L}^{\beta_{2},(s)}_{13} \mathrm{L}^{\beta_{3},(s)}_{12}; \quad \alpha_{1,2,3}=\{0,1\},\label{tza:TZA}
\end{align}
generalizing the above two relations \eqref{tza:L000} and \eqref{tza:L011}. It was introduced by Korepanov in \cite{Korepanov1993,Korepanov1994c,Korepanov1994,Korepanov2013} to investigate Zamolodchikov' tetrahedron equation, whcih underlines the symmetry of a three-dimensional integrable model, generalizing the relation of the Yang-Baxter equation to two-dimensional integrable models. As was shown in \cite{Shiroishi1995as,Shiroishi1995vc,Umeno1998b} the tetrahedral Zamolodchikov algebra can be used in order to construct an interacting model of spin $s=\nicefrac{1}{2}$ fermions, and, in particular, to obtain the one-dimensional Hubbard model (see \cite{Essler:2005bk} for a review, and there references therein).

To obtain the coefficients $S^{\alpha_{1} \alpha_{2} \alpha_{3}}_{\beta_{1}\beta_{2}\beta_{3}}$  in \eqref{tza:TZA} one has to evaluate generic tensor products $ \mathrm{L}^{\alpha_{1},(s)}_{12} \mathrm{L}^{\alpha_{1},(s)}_{13} \mathrm{L}^{\alpha_{3},(s)}_{23}$ and \newline $\mathrm{L}^{\beta_{1},(s)}_{23} \mathrm{L}^{\beta_{1},(s)}_{13} \mathrm{L}^{\beta_{3},(s)}_{12}$ that appear in the left and right hand sides of \eqref{tza:TZA}. To this end, we note that both $\mathrm{L}^{0,(s)}_{jk}$ and $\mathrm{L}^{1,(s)}_{jk}$ can be written in the following general form:
\begin{align}
    \Gamma_{jk}(u;\zeta_{j},\zeta_{k}) &=h_{0}(u;\zeta_{j},\zeta_{k})+h_{1}(u;\zeta_{j},\zeta_{k})n_{j} +h_{2}(u;\zeta_{j},\zeta_{k})n_{k}
    +h_{3}(u;\zeta_{j},\zeta_{k})n_{j}n_{k} +h_{4}(u;\zeta_{j},\zeta_{k})\Delta_{jk} \nonumber \\ &+h_{5}(u;\zeta_{j},\zeta_{k})\Delta_{kj} +h_{6}(u;\zeta_{j},\zeta_{k})\tilde{\Delta}^{(\dagger)}_{jk}+h_{7}(u;\zeta_{j},\zeta_{k})\tilde{\Delta}_{jk} ,\label{bs:fermionic_R_generic_form}
\end{align}
One then obtains either $\mathrm{L}^{0,(s)}_{jk}$ or $\mathrm{L}^{1,(s)}_{jk}$ by fixing accordingly the functions $\left(h_{0}(u;\zeta_{j},\zeta_{k}),\ldots,h_{7}(u;\zeta_{j},\zeta_{k}\right)$ in terms of the coefficients of the $R$-operator \eqref{bs:a_a_prime}-\eqref{bs:d_d_prime}. Thus, one can evaluate the general tensor products  $\Gamma_{12}(w;\zeta_{1},\zeta_{2})\,\Gamma_{13}(u;\zeta_{1},\zeta_{3})\,\Gamma_{23}(v;\zeta_{2},\zeta_{3})$ and $\Gamma_{23}(v';\zeta'_{2},\zeta'_{3})\,\Gamma_{13}(u';\zeta'_{1},\zeta'_{3})\,\Gamma_{12}(w';\zeta'_{1},\zeta'_{2})$ to take into account all possible permutations of indices $\alpha_{1,2,3}$ and $\beta_{1,2,3}$ in \eqref{tza:TZA}. The result of this very lengthy computation is a set of 64 algebraic equations that will be presented elsewhere. Below we give a particular solution to this set of equations.

\subsection{Solving the tetrahedral Zamolodchikov's algebra}
In what follows we omit the superscript $(s)$ everywhere to simplify the notations, and consider a particular solution corresponding to the case of $k=0,\, \zeta_{i}=\zeta=\nicefrac{\pi}{2}$. It follows from the Hamiltonian \eqref{bs:XY_Hamiltonian} that this case corresponds to the fermionic $XX$-model. To further simplify our notations we also denote:
\begin{align}
    \Lambda^{\alpha_{1}\alpha_{2}\alpha_{3}} \equiv \mathrm{L}^{\alpha_{1}}_{12} \mathrm{L}^{\alpha_{2}}_{13} \mathrm{L}^{\alpha_{3}}_{23}; \quad  \tilde{\Lambda}^{\beta_{1}\beta_{2}\beta_{3}} \equiv \mathrm{L}^{\beta_{1}}_{23} \mathrm{L}^{\beta_{2}}_{13} \mathrm{L}^{\beta_{3}}_{12}.\label{tza:products_notation}
\end{align}
Our main result is a solution to the tetrahedral Zamolodchikov algebraic relations, in addition to those in \eqref{tza:L000} and \eqref{tza:L011}, with the following non-trivial coefficients $S^{\alpha_{1} \alpha_{2} \alpha_{3}}_{\beta_{1}\beta_{2}\beta_{3}}$:\footnote{We note here, that the solution is not unique, and can be represented in a number of forms.}
\begin{align}
    \Lambda^{110} &=\tilde{\Lambda}^{011}\label{tza:L110},\\
    \Lambda^{101} &=\tilde{\Lambda}^{101}\label{tza:L101},\\
    \Lambda^{111} &=\left(S^{111}_{010}\right)\tilde{\Lambda}^{010}+\left(S^{111}_{100}\right)\tilde{\Lambda}^{100}+\left(S^{111}_{110}\right)\tilde{\Lambda}^{110}+\left(S^{111}_{111}\right)\tilde{\Lambda}^{111}, \label{tza:L111} \\
    \Lambda^{001} &=\left(S^{001}_{010}\right)\tilde{\Lambda}^{010}+\left(S^{001}_{100}\right)\tilde{\Lambda}^{100}+\left(S^{001}_{110}\right)\tilde{\Lambda}^{110}+\left(S^{001}_{111}\right)\tilde{\Lambda}^{111}, \label{tza:L001}\\
    \Lambda^{010} &=\left(S^{010}_{010}\right)\tilde{\Lambda}^{010}+\left(S^{010}_{100}\right)\tilde{\Lambda}^{100}+\left(S^{010}_{110}\right)\tilde{\Lambda}^{110}+\left(S^{010}_{111}\right)\tilde{\Lambda}^{111}, \label{tza:L010}\\
    \Lambda^{100} &=\left(S^{100}_{001}\right)\tilde{\Lambda}^{001}+\left(S^{100}_{100}\right)\tilde{\Lambda}^{100}+\left(S^{100}_{101}\right)\tilde{\Lambda}^{101}+\left(S^{100}_{111}\right)\tilde{\Lambda}^{111}, \label{tza:L100}
\end{align}
where:
\begin{equation*}
\begin{aligned}[t]
       S^{111}_{010}&=-\cos\left(\frac{\eta_{23}}{2}\right)\sec\left(\frac{\eta_{12}}{2}\right)\sec\left(\frac{\eta_{13}}{2}\right),\nonumber  \\ 
       S^{111}_{100}&=\sin\left(\frac{\eta_{13}}{2}\right)\sec\left(\frac{\eta_{12}}{2}\right)\csc\left(\frac{\eta_{23}}{2}\right),
       \nonumber\\
       S^{111}_{110}&= 2 \mi \tan\left(\frac{\eta_{12}}{2}\right)\tan\left(\frac{\eta_{13}}{2}\right)\cot\left(\frac{\eta_{23}}{2}\right),
       \nonumber \\
       S^{111}_{111}&=\tan\left(\frac{\eta_{13}}{2}\right)\cot\left(\frac{\eta_{23}}{2}\right), 
       \nonumber\\
        S^{001}_{010}&= \cos\left(\frac{\eta_{23}}{2}\right)\sec\left(\frac{\eta_{12}}{2}\right)\sec\left(\frac{\eta_{13}}{2}\right), \nonumber \\ 
       S^{001}_{100}&=-\tan\left(\frac{\eta_{12}}{2}\right)\cot\left(\frac{\eta_{23}}{2}\right),
       \nonumber\\
       S^{001}_{110}&=-2 \mi \tan\left(\frac{\eta_{12}}{2}\right)\tan\left(\frac{\eta_{13}}{2}\right)\cot\left(\frac{\eta_{23}}{2}\right),
       \nonumber \\
       S^{001}_{111}&=-\sin\left(\frac{\eta_{12}}{2}\right)\csc\left(\frac{\eta_{23}}{2}\right)\sec\left(\frac{\eta_{13}}{2}\right), 
       \nonumber
    \end{aligned} \qquad \qquad 
    \begin{aligned}[t]
    S^{010}_{010}&=-\tan\left(\frac{\eta_{12}}{2}\right)\tan\left(\frac{\eta_{13}}{2}\right), \nonumber \\ 
    S^{010}_{100}&=\csc\left(\frac{\eta_{23}}{2}\right)\sec\left(\frac{\eta_{12}}{2}\right)\sin\left(\frac{\eta_{13}}{2}\right),
         \nonumber \\
    S^{010}_{110}&=2 \mi \cot\left(\frac{\eta_{23}}{2}\right)\tan\left(\frac{\eta_{12}}{2}\right)\tan\left(\frac{\eta_{13}}{2}\right), 
       \nonumber\\
    S^{010}_{111}&=\csc\left(\frac{\eta_{23}}{2}\right)\sec\left(\frac{\eta_{13}}{2}\right)\sin\left(\frac{\eta_{12}}{2}\right), \nonumber \\ 
    S^{100}_{001}&=-\cot\left(\frac{\eta_{12}}{2}\right)\cot\left(\frac{\eta_{13}}{2}\right),\nonumber \\
    S^{100}_{100}&=-\cos\left(\frac{\eta_{12}}{2}\right)\csc\left(\frac{\eta_{13}}{2}\right)\csc\left(\frac{\eta_{23}}{2}\right), 
       \nonumber\\
    S^{100}_{101}&=4 \mi\frac{\sin(\eta_{12})\sin(\eta_{13})\sin(\eta_{23})}{\left(\sin(\eta_{12})+\sin(\eta_{13})+\sin(\eta_{23})\right)^{2}}, 
       \nonumber\\
    S^{100}_{111}&=-\cos\left(\frac{\eta_{13}}{2}\right)\csc\left(\frac{\eta_{12}}{2}\right)\csc\left(\frac{\eta_{23}}{2}\right), 
       \nonumber
\end{aligned}                             
\end{equation*}
In addition, we note that the set of operators $\Lambda^{\alpha_{1}\alpha_{2}\alpha_{3}}$ and $\tilde{\Lambda}^{\beta_{1}\beta_{2}\beta_{3}}$ are linearly dependent. One can check, for example, the following relation between $\tilde{\Lambda}^{\alpha_{1}\alpha_{2}\alpha_{3}}$:
\begin{align}
    \tilde{\Lambda}^{111}&=\left(\xi^{111}_{110}\right)\tilde{\Lambda}^{110}+\left(\xi^{111}_{101}\right)\tilde{\Lambda}^{101}+\left(\xi^{111}_{001}\right)\tilde{\Lambda}^{001}
    +\left(\xi^{111}_{010}\right)\tilde{\Lambda}^{010}+\left(\xi^{111}_{100}\right)\tilde{\Lambda}^{100},
\end{align}
where the linear coefficients have the form:
\begin{equation*}
\begin{aligned}[t]
       \xi^{000}_{110}&=4 \mi \csc\left(\eta_{12}\right)\sec\left( \frac{\eta_{12}+\eta_{13}}{2}\right)\sin^{3}\left(\frac{\eta_{12}}{2}\right)\sin\left(\frac{\eta_{13}}{2}\right), \nonumber \\
 \xi^{000}_{101}&=2 \mi \cos\left(\frac{\eta_{12}}{2}\right)\cos\left(\frac{\eta_{13}}{2}\right)\cot\left(\frac{\eta_{13}}{2}\right)\sec\left(\frac{\eta_{12}+\eta_{13}}{2}\right), \nonumber \\
 \xi^{000}_{001}&=\cot\left(\frac{\eta_{13}}{2}\right)\sec\left(\frac{\eta_{12}+\eta_{13}}{2}\right)\sin\left(\frac{\eta_{23}}{2}\right),\nonumber \\
  \xi^{000}_{010}&=\sec\left(\frac{\eta_{12}+\eta_{13}}{2}\right)\sin\left(\frac{\eta_{23}}{2}\right)\tan\left(\frac{\eta_{12}}{2}\right),\nonumber \\
   \xi^{000}_{100}&=\cot\left(\frac{\eta_{13}}{2}\right)\tan\left(\frac{\eta_{12}}{2}\right).
    \end{aligned}                            
\end{equation*}

We therefore have found a solution which depends only on the differences of the spectral parameters $\eta_{ij}=u_{j}-u_{k}$. This is of course the consequence of the decorated Yang-Baxter equations \eqref{tza:DYBE} having the same dependence on $\eta_{ij}$, and is in contrast to the solution of the tetrahedral Zamolodchikov algebra given in \cite{Shiroishi1995as,Shiroishi1995vc,Umeno1998b} where the solution for the coefficients $S^{\alpha_{1} \alpha_{2} \alpha_{3}}_{\beta_{1}\beta_{2}\beta_{3}}$ is not of the difference type. As we explained in the introduction, the solution that depends on the differences of the spectral parameters is natural for taking the continuous limit and obtaining an integrable model of relativistic fermions.

\section{Conclusion}
The solution of the tetrahedral Zamolodchikov algebra given in the previous section can be used, as shown in \cite{Essler:2005bk,Shiroishi1995as,Shiroishi1995vc,Umeno1998b} to obtain, in principle, an interacting model from the free fermion model described by the $\mathcal{R}$-operator \eqref{bs:fermionic_R_spin}. To this end, one defines an $R$-matrix of the form:
\begin{align}
   \tilde{\mathcal{R}}_{jk}(\eta_{jk})=\mathrm{L}^{0,(\uparrow)}_{jk}\otimes\mathrm{L}^{0,(\downarrow)}_{jk}+\alpha_{jk}\mathrm{L}^{0,(\uparrow)}_{jk}\otimes\mathrm{L}^{1,(\downarrow)}_{jk}+\beta_{jk}\mathrm{L}^{1,(\uparrow)}_{jk}\otimes\mathrm{L}^{0,(\downarrow)}_{jk}+\gamma_{jk}\mathrm{L}^{1,(\uparrow)}_{jk}\otimes\mathrm{L}^{1,(\downarrow)}_{jk}, \label{con:new_R}
\end{align}
where the coefficients $\alpha_{jk},\beta_{jk},\gamma_{jk}$ are to be determined from the condition that the $\mathbb{R}_{jk}$-operator \eqref{con:new_R} satisfies the Yang-Baxter equation. This will be investigated in the future work. 

It will be also interesting to obtain a more general solutions than the specific one considered in this paper. This may help to gain further insights on passing from lattice to continuous models, and explain the appearance of the non-ultralocal structures \eqref{intro:nonultralocal}, such as the one appearing from string theory \eqref{intro:aaf_lag}, and, therefore, to find new methods to quantize such non-ultralocal integrable systems, which remains an open problem.

\appendix
\section{Appendix}
\label{appendix}
In this appendix we list some useful identities and formulas used to calculate the tetrahedral Zamolodchikov algebra. Starting from the fermionic $R$-operator \eqref{bs:fermionic_R}:
\begin{align}
    R_{jk}(u;\zeta_{j},\zeta_{k})&=a(u;\zeta_{j},\zeta_{k})\left[-n_{j} n_{k} \right] +a(u;\zeta_{j},\zeta_{k})\left[(1-n_{j})(1- n_{k}) \right]+b(u;\zeta_{j},\zeta_{k})\left[n_{j}(1- n_{k}) \right] \nonumber\\
    &+b'(u;\zeta_{j},\zeta_{k})\left[n_{k}(1- n_{j})\right] +c(u;\zeta_{j},\zeta_{k})\left[\Delta_{jk}+\Delta_{kj} \right] +d(u;\zeta_{j},\zeta_{k})\left[-\tilde{\Delta}^{(\dagger)}_{jk}-\tilde{\Delta}_{jk}\right] ,\label{app:fermionic_R}
\end{align}
one can show the following identities used to derive the relation \eqref{tza:identity_main}, as well as the decorated Yang-Baxter equation \eqref{tza:DYBE}:
\begin{align}
    (2n_{j}-1)R_{jk}(u;\zeta_{j},\zeta_{k})  (2n_{k}-1) &=  (2n_{k}-1)R_{jk}(u;\zeta_{j},\zeta_{k})(2n_{j}-1),\\
     (2n_{j}-1)R_{jk}(u;\zeta_{j},\zeta_{k})  (2n_{j}-1) &=  (2n_{k}-1)R_{jk}(u;\zeta_{j},\zeta_{k})(2n_{k}-1).
\end{align}
More explicitly, one has the following relations:
\begin{align}
    R_{jk}(u;\zeta_{j},\zeta_{k})(2n_{j}-1)&=a(u;\zeta_{j},\zeta_{k})\left[-n_{j} n_{k} \right] -a(u;\zeta_{j},\zeta_{k})\left[(1-n_{j})(1- n_{k}) \right] \nonumber\\
   & +b(u;\zeta_{j},\zeta_{k})\left[n_{j}(1- n_{k}) \right]
    -b'(u;\zeta_{j},\zeta_{k})\left[n_{k}(1- n_{j})\right]\nonumber\\
     &+c(u;\zeta_{j},\zeta_{k})\left[-\Delta_{jk}+\Delta_{kj} \right] +d(u;\zeta_{j},\zeta_{k})\left[\tilde{\Delta}^{(\dagger)}_{jk}-\tilde{\Delta}_{jk}\right],\label{app:fermionic_R_id1}\\
    \nonumber \\
   (2n_{j}-1) R_{jk}(u;\zeta_{j},\zeta_{k})&=a(u;\zeta_{j},\zeta_{k})\left[-n_{j} n_{k} \right] -a(u;\zeta_{j},\zeta_{k})\left[(1-n_{j})(1- n_{k}) \right]\nonumber\\
   &+b(u;\zeta_{j},\zeta_{k})\left[n_{j}(1- n_{k}) \right] 
    -b'(u;\zeta_{j},\zeta_{k})\left[n_{k}(1- n_{j})\right]\nonumber\\
    & +c(u;\zeta_{j},\zeta_{k})\left[\Delta_{jk} - \Delta_{kj} \right] +d(u;\zeta_{j},\zeta_{k})\left[-\tilde{\Delta}^{(\dagger)}_{jk}+\tilde{\Delta}_{jk}\right],
    \label{app:fermionic_R_id2}
\end{align}
\bibliographystyle{elsarticle-num}

\bibliography{bs_zam}

\end{document}